\documentclass[12pt]{iopart}
\usepackage{harvard}
\usepackage{bm}
\usepackage{graphicx}
\newcommand{\EQ}{\begin{equation}}
\newcommand{\EN}{\end{equation}}
\newcommand{\EQA}{\begin{eqnarray}}
\newcommand{\ENA}{\end{eqnarray}}

\newcommand{\Eq}[1]{equation~(\ref{#1})}

\newcommand{\Fig}[1]{Fig.~\ref{#1}}

\newcommand{\bra}[1]{\langle #1\rangle}

{}
{}

{}
{}
{}
{}
{}
{}
{}
{}
{}
{}
{}
\newcommand{\meanUU}{\overline{\mbox{\boldmath $U$}}{}}{}
{}
{}

\newcommand{\meanU}{\overline{U}}

{}
{}
{}
{}

\newcommand{\ppom}{\bm{\hat{\varpi}}}
\newcommand{\uu}{\mbox{\boldmath $u$} {}}
\newcommand{\UU}{\mbox{\boldmath $U$} {}}
\newcommand{\xx}{\mbox{\boldmath $x$} {}}

\newcommand{\bb}{\mbox{\boldmath $b$} {}}
\newcommand{\BB}{\mbox{\boldmath $B$} {}}

\newcommand{\FF}{\mbox{\boldmath $F$} {}}

\newcommand{\grav}{\mbox{\boldmath $g$} {}}
\newcommand{\nab}{\mbox{\boldmath $\nabla$} {}}
\newcommand{\OO}{\mbox{\boldmath $\Omega$} {}}
\newcommand{\oo}{\mbox{\boldmath $\omega$} {}}

\newcommand{\dd}{{\rm d} {}}

\def\Rey{\mbox{\rm Re}}

\def\half{{\textstyle{1\over2}}}
\def\threehalf{{\textstyle{3\over2}}}

\def\onethird{{\textstyle{1\over3}}}

\newcommand{\K}{\,{\rm K}}
\newcommand{\g}{\,{\rm g}}

\newcommand{\cmcube}{\,{\rm cm^{-3}}}
\newcommand{\m}{\,{\rm m}}

\newcommand{\kms}{\,{\rm km/s}}

\newcommand{\AU}{\,{\rm AU}}

\newcommand{\yapj}[3]{ #1, {ApJ,} {#2}, #3}

\newcommand{\yapjl}[3]{ #1, {ApJ,} {#2}, #3}

\newcommand{\yana}[3]{ #1, {A\&A,} {#2}, #3}

\newcommand{\ypasj}[3]{ #1, {Publ.\ Astron.\ Soc.\ Japan,} {#2}, #3}

\newcommand{\yprl}[3]{ #1, {Phys.\ Rev.\ Lett.,} {#2}, #3}

\newcommand{\ymn}[3]{ #1, {MNRAS,} {#2}, #3}
\newcommand{\ynat}[3]{ #1, {Nature,} {#2}, #3}

\newcommand{\yicarus}[3]{ #1, {Icarus,} {#2}, #3}

\newcommand{\yjour}[4]{ #1, {#2}, {#3}, #4}

\newcommand{\ybook}[3]{ #1, {#2} (#3)}
\newcommand{\yproc}[5]{ #1, in {#3}, ed.\ #4 (#5), #2}

\usepackage{iopams}  

\begin{document}

\title[Turbulent protostellar discs]{Turbulent protostellar discs}

\author{A Brandenburg}

\address{NORDITA, Roslagstullsbacken 23, SE-10691 Stockholm, Sweden}

\ead{brandenb@nordita.org}

\begin{abstract}
Aspects of turbulence in protostellar accretion discs are being reviewed.
The emergence of dead zones due to poor ionization and alternatives
to the magneto-rotational instability are discussed.
The coupling between dust and gas in protostellar accretion discs is
explained and turbulent drag is compared with laminar drag in the
Stokes and Epstein regimes.
Finally, the significance of magnetic field generation in turbulent
discs is emphasized in connection with driving outflows and with
star-disc coupling.
\end{abstract}

% Example for Physics and Astronomy Classification System (PACS) codes.
% For a list of GEOPHYSICS, ASTRONOMY, AND ASTROPHYSICS PACS codes see
% http://www.aip.org/pacs/pacs06/pacs0690.html

% The command \maketitle forces a page break after the point where it is inserted and so to keep the header material on a separate page from the body of the text insert \maketitle or \newpage before the start of the text. If \maketitle is not included the text of the article will start immediately after the abstract.
% \maketitle

\section{Introduction} 

One of the amazing properties of accretion disc theory is the fact
that it is applicable over a vast range of scales, from discs around
supermassive black holes to those around stellar mass black holes
and around neutron stars, to young stellar objects and possibly
even protoplanets.
This is mainly because the equations of hydrodynamics and
magnetohydrodynamics are scale-invariant over broad ranges.
In all cases accretion is caused by the action of the Reynolds and
Maxwell stresses.
The stress is assumed to be expressible in terms of the radial angular
velocity gradient multiplied by some turbulent viscosity coefficient.
However, a major problem is that in protostellar discs the temperatures
are rather low, so the degree of ionization depends significantly on
cosmic ray ionization.
At a radial distance of about $1\AU$ the regions near the
midplane are strongly shielded, which implies the possibility of
``dead zones'' (Gammie 1996) where magnetic effects cannot play a role.
This leads to our first topic.

Accretion is necessary for accumulating the mass of the central
object which is in this case a protostar.
This accretion is presumably of turbulent origin, because the
molecular viscosity is obviously far too small.
Nevertheless, the molecular or microscopic viscosity together with
microscopic magnetic diffusion are crucial, and they are in fact the
main processes that can heat the disc--in addition to irradiation
from the central star onto the our surface of the disc.
However, it is still unclear whether in a turbulent disc planet formation
occurs because of turbulence or in spite of it.
This will be our second topic.

Finally, we address the issue of outflows from the star and from the disc.
The occurrence of collimated outflows in star-forming regions has always
been associated with magnetic fields.
Given that magnetic fields can also be produced within the disc from
which such outflows emanate, the question emerges whether the field
necessary for collimation has to be an external interstellar field,
or whether the disc field suffices.
In addition, there will be the stellar field and its coupling to
the ambient magnetic disc.
This is thought to be critical for spinning
up the protostar during its formation and spinning it down
later during its evolution.
This is the final topic discussed in this review.
We start however with a discussion of the mean accretion stress,
its definition, and how it is used.

\section{The mean accretion stress}

In order to get an idea about the nature of turbulence in accretion discs,
it is useful to consider simulations in a local cartesian geometry
with shearing sheet boundary conditions in the radial direction.
An important output parameter of such simulations is the Shakura--Sunyaev
viscosity parameter, $\alpha_{\rm SS}$, which is a non-dimensional measure
of the turbulent viscosity $\nu_{\rm t}$, in terms of the
sound speed $c_{\rm s}$ and the disc scale height $H$, i.e.\
\EQ
\nu_{\rm t}=\alpha_{\rm SS}c_{\rm s}H.
\label{nut}
\EN
This coefficient is used analogously to the molecular viscosity and hence
is used to multiply the velocity gradient to get the accretion stress.
The main difference is however that the turbulent viscosity multiplies
the gradient of the {\it mean} velocity while the actual viscosity multiplies
the actual turbulent velocity gradients, which can be much higher locally
on very small scales.
In the mean field description the mean stress is written as
\EQ
\overline{\tau}_{\varpi\phi}=
-\rho\nu_{\rm t}\varpi{\partial\overline{\Omega}\over\partial\varpi},
\label{stressMF}
\EN
where $\varpi$ is the cylindrical radius, $\rho$ is the density,
$\overline{\Omega}$ is the mean angular velocity, and
$\overline{\tau}_{\varpi\phi}$ is the total `horizontal' stress.
This component of the stress tensor enters the conservation equation
for the mean angular momentum,
\EQ
{\partial\over\partial t}\left(\rho\varpi^2\overline{\Omega}\right)
+\nab\cdot\left(\rho\,\meanUU\,\meanU_\phi
+\ppom\overline{\tau}_{\varpi\phi}\right)=0.
\EN
Here, $\ppom=(1,0,0)^{\rm T}$ is the unit vector in the direction away
from the axis, and overbars denote mean quantities that are obtained by
averaging over the azimuthal direction.
The velocity is then written as $\UU=\meanUU+\uu$, and fluctuations
of the density $\rho$ are ignored.
Also, for simplicity, we have ignored the vertical component of the stress,
$\overline{\tau}_{z\phi}$.
Roughly, the two terms under the divergence balance, so a positive
stress, $\overline{\tau}_{\varpi\phi}>0$, causes a negative (inward)
accretion velocity, $\meanU_\varpi<0$.

\begin{figure}[t!]
\resizebox{\hsize}{!}{\includegraphics[clip=true]{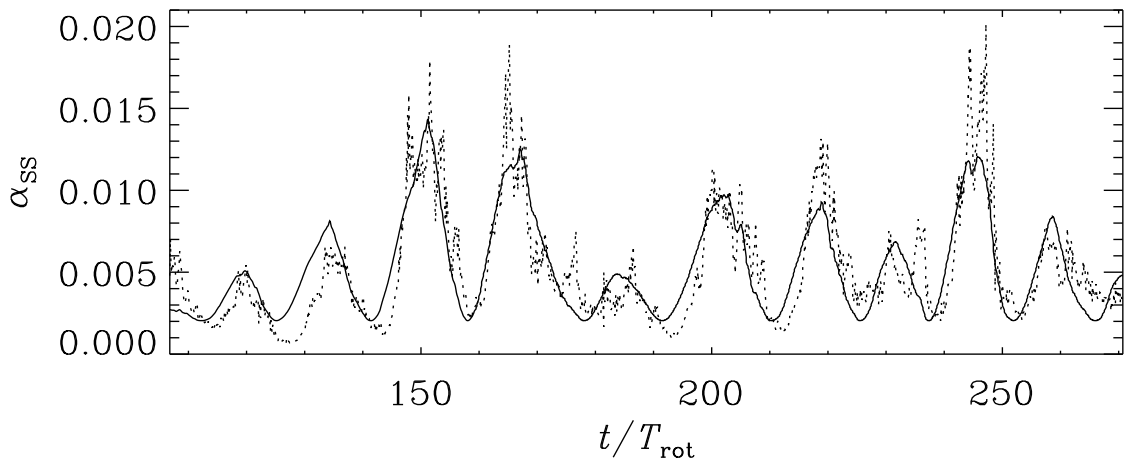}}
\caption{\footnotesize
Time series of the Shakura-Sunyaev viscosity alpha, $\alpha_{\rm SS}(t)$
(dotted line), compared with a fit of the form \Eq{alpfit} in terms
of the instantaneous magnetic field strength (solid line).
Time is normalized in terms of the orbital time,
$T_{\rm rot}=2\pi/\Omega$, where $\Omega$ is the local
angular velocity.
Note that $\alpha_{\rm SS}$ fluctuates strongly in time
about an average value of around 0.007.
[Adapted from Brandenburg (1998).]
}
\label{alpt}
\end{figure}

In a theory that works with mean velocities, $\meanUU$,
as opposed to the actual one, $\UU$, one needs to calculate the
contributions of small scale velocities and magnetic fields on the
mean flow through the sum of mean Reynolds and Maxwell stresses,
\EQ
\overline{\tau}_{\varpi\phi}=-\overline{b_\varpi b_\phi}/\mu_0
+\rho \overline{u_\varpi u_\phi},
\label{stress}
\EN
where $\bb$ and $\uu$ denote the fluctuating components of the
magnetic and velocity fields, and $\mu_0$ is the vacuum permeability.
The first term, i.e.\ the magnetic stress, is usually the largest term.
The time dependence of the resulting stress determined from a
simulation, and expressed in nondimensional form as
\EQ
\alpha_{\rm SS}(t)
=\overline{\tau}_{\varpi\phi}/\left(\rho c_{\rm s}H\threehalf\Omega\right),
\EN
is shown in \Fig{alpt}.
The simulation domain covers only the upper disc plane.
The instantaneous time dependence follows closely that of the
instantaneous magnetic field strength and can be described by
a fit of the form
\EQ
\alpha_{\rm SS}(t)\approx\alpha_{\rm SS}^{(0)}
+\alpha_{\rm SS}^{(B)}{\bra{\BB}^2\over B_0^2},
\label{alpfit}
\EN
where $\bra{\BB}$ is the mean magnetic field averaged over the entire
upper disc plane, and $B_0=\sqrt{\mu_0\rho}c_{\rm s}$ is the thermal
equipartition field strength, with $c_{\rm s}$ being the sound speed,
and fit parameters $\alpha_{\rm SS}^{(0)}\approx0.002$ and
$\alpha_{\rm SS}^{(B)}\approx0.06$ (Brandenburg 1998).
Blackman et al.\ (2008) have shown that a similar relationship,
albeit without the $\alpha_{\rm SS}^{(0)}$ term, can be
established for a number of other simulations published in the literature.

Simulations based on the magneto-rotational instability
(Balbus \& Hawley 1991, 1998) show that
the average value of $\alpha_{\rm SS}(t)$ is small (around 0.007;
see Hawley et al.\ 1995, 1996, Stone et al.\ 1996,
Brandenburg et al.\ 1995, 1996).
Larger values of $\alpha_{\rm SS}$ are obtained with an externally
imposed vertical or axial magnetic field
(e.g.\ Hawley et al.\ 1995, Torkelsson et al.\ 1996),
as is also predicted from \Eq{alpfit}.
However, $\alpha_{\rm SS}$ may well be smaller, depending on simulation
details.\footnote{Here the scale height has been defined such that
$\rho\sim\exp(-z^2/H^2)$.
Later we shall define a scale height such that
$\rho\sim\exp(-\half z^2/\tilde{H}^2)$.
Note that if $\tilde{H}=H/\sqrt{2}$ were used in the definition of
$\alpha_{\rm SS}$ in \Eq{nut}, the resulting value of $\alpha_{\rm SS}$
would be larger by a factor $\sqrt{2}$.
Note also that in some definitions of $\alpha_{\rm SS}$ the factor
$3/2$ from $\varpi\partial\Omega/\partial\varpi=-\threehalf\Omega$
is neglected, in which case $\alpha_{\rm SS}$ would be larger by
yet another factor of 3/2, so altogether $\threehalf\sqrt{2}\approx2.1$.}
Indeed, there is still an open issue regarding the convergence of the
numerical results (see, e.g., Fromang \& Papaloizou 2007, Fromang et al.\ 2007).

The variation of $\alpha_{\rm SS}(t)$ shows a typical
time scale of around 15 orbits.
In a simulation with vertical density stratification this can be
a consequence of the emergence of cyclic large scale dynamo action.
The magnetic field reverses between two maxima, so the actual period
is around 30 orbits.
Physically, this time scale is related to the turbulent diffusion time.
This is now reasonably well understood and the details of the large-scale
dynamo magnetic field depend obviously on boundary conditions.
Such cycles are not expected in global geometry, as can be demonstrated
by corresponding mean field calculations (Bardou et al.\ 2001).

Global simulations of accretion disc turbulence have been studied by
various groups since the late 1990s.
A conceptionally simple approach is that of the cylindrical disc
which omits vertical stratification.
In such simulations the magnetic field is able to develop over
large scales and can be substantially stronger than in local simulations,
resulting therefore also in larger values of $\alpha_{\rm SS}$;
Armitage (1998) found $\alpha_{\rm SS}\approx0.1$.
Similar results where later confirmed with fully global simulations
by Hawley (2000), who found values of  $\alpha_{\rm SS}$ in the range
of 0.1--0.2.
Simulations of global discs permeated by a magnetic field from a
central star also obey \Eq{alpfit}, but in the simulations of
Steinacker \& Papaloizou (2002) the magnetic field was weaker,
resulting in $\alpha_{\rm SS}$ values of around 0.004.
Similar figures have also been reported by Nelson \& Papaloizou (2003)
and Fromang \& Nelson (2006) in the context of turbulent protoplanetary discs
with and without planet--disc interaction.
Global simulations have now also been able to produce estimates for
Reynolds and Maxwell stresses as a function of radius using the
disc-in-a-box approach (Lyra et al.\ 2008).
Among other things they find that larger Mach numbers result in
larger normalized accretion stresses, as quantified by the
Shakura--Sunyaev viscosity alpha.

\section{Dead zones}

When the gas is cool, the main source of ionization is not thermal,
but it is cosmic ray ionization from the galaxy and UV ionization from
the central star (and possibly other nearby stars).
Research in this direction has been pursued by Glassgold et al.\ (1997)
and Igea et al.\ (1999).
Cosmic ray ionization usually plays an important role at moderate and large
distances from the central object and away from the midplane where
cosmic rays are shielded.
Here, the column density of the gas is so high that the degree of
ionization is very small, and the magneto-rotational instability (MRI)
cannot grow because of the large diffusivity of the field.
In addition, magnetic fields and a magnetized wind from the central star
have a tendency to shield the disc from cosmic rays.
The resulting region of low ionization in the disc
is therefore referred to as the dead zone (Gammie 1996).

Recent work of Fleming \& Stone (2003) shows that, although the local
Maxwell stress drops to negligible values in the dead zones, the
Reynolds stress remains approximately independent of height and never
drops below approximately 10\% of the maximum Maxwell stress,
provided the column density in that zone is less that 10 times
the column density of the active layers.
The non-dimensional ratio of stress to gas pressure is just
the Shakura \& Sunyaev (1973) viscosity parameter, $\alpha_{\rm SS}$.
Fleming \& Stone (2003) find typical values of a few times $10^{-4}$ in
the dead zones and a few times $10^{-3}$ in the MRI-active layers.
Similar results have recently been confirmed by Oishi et al.\ (2007).
On the other hand, Inutsuka \& Sano (2005) have questioned the very
existence of dead zones, and proposed that the turbulent dissipation in
the disc provides sufficient energy for the ionization.
However, their calculation assumes a magnetic Reynolds
number that is much smaller than expected from simulations
(see Matsumura \& Pudritz 2006).

The radial extent of the dead zone depends primarily on
the disc column density
as well as effects of grains and radiation chemistry
(e.g., Sano et al.\ 2000, Matsumura \& Pudritz 2005).
Inside the dead zone the magnetic Reynolds number tends to be
below a certain critical value that is somewhere between 1 and 100
(Sano \& Stone 2002), making MRI-driven turbulence impossible.
Estimates for the radial extent of the dead zone range from
0.7--100 AU (Fromang et al.\ 2002),
to 2--20 AU in calculations by Semenov et al.\ (2004).
For smaller radii, thermal and UV ionization are mainly responsible for
sustaining some degree of ionization (Glassgold et al.\ 1997;
Igea et al.\ 1999).
The significance of the reduced value of $\alpha_{\rm SS}$ in the dead zones
is that it provides a mechanism for stopping the inward migration of
Jupiter-sized planets (Matsumura \& Pudritz 2005, 2006).
Jets can still be launched from the well-coupled surface layer
above the dead zone (e.g.\ Li 1996; Campbell 2000).

Although the MRI may be inactive in the bulk of the disc,
there are still alternative mechanisms of angular momentum transport.
In protostellar discs there are probably at least two other mechanisms
that might contribute to the accretion torque:
density waves (R\'o\.zyczka \& Spruit 1993)
and the interaction with other planets in the disc
(Goodman \& Rafikov 2001).
An additional mechanisms that has been investigated recently is the so-called
streaming instability that results from the dust trying to move at
Keplerian speed relative to the surrounding gas that rotates
slightly sub-Keplerian due to partial pressure support
(Youdin \& Johansen 2007, Johansen \& Youdin 2007).

Another frequently discussed proposal is the possibility that the
turbulence in the disc might be driven by a nonlinear
finite amplitude instability (Chagelishvili et al.\ 2003).
While Balbus et al.\ (1996) and Hawley et al.\ (1999) have given general
arguments against this possibility, Afshordi et al.\ (2005) and other
groups have continued to investigate the so-called bypass to turbulence.
The basic idea is that successive strong transients can maintain a
turbulent state in a continuously excited manner.
Lesur \& Longaretti (2005) have recently been able to quantify more
precisely the critical Reynolds number required for instability.
They have also highlighted the importance of pressure fluctuations that
demonstrate that a general argument by Balbus et al.\ (1996)
against nonlinear hydrodynamic instabilities is insufficient.

Yet another possibility is to invoke convectively stable vertical density
stratification which may give rise to the so-called strato-rotational
instability (Dubrulle et al.\ 2005).
This instability was recently discovered by Molemaker et al.\ (2001) and
the linear stability regime was analyzed by Shalybkov \& R\"udiger (2005).
However, the presence of no-slip radial boundary conditions that are
relevant to experiments are vital for the mechanism to work.
Indeed, the instability vanishes for an unbounded regime, making it
irrelevant for accretion discs (Umurhan 2006).

\section{Dust dynamics in disc turbulence}

An important aspect of protostellar discs is their capability to
produce planets.
This requires the local accumulation of dust to produce larger
conglomerates that can condense into rocks and boulders that
are big enough to be decoupled from their ambient gas flow, and
that can combine to grow under the influence of self-gravity.

The dynamics of dust particles depends critically on their size.
Small particles are essentially advected by the gas flow, whereas larger
ones do not significantly interact viscously with the gas and their
dynamics is essentially governed by the gravitational field leading to
Keplerian orbital motion and to settling toward the midplane.
If the dust grains are small, turbulence will stir up the dust,
preventing it from settling.
Particles of radius $a_{\rm p}$ can be considered big if its so-called
stopping time, $\tau_{\rm s}$, is long compared with the orbital time.
The stopping time is essentially the time it takes for a particle
to decelerate by a factor of $1/e$.
The precise expression for $\tau_{\rm s}$ depends critically on the
ratio of the radius of the particle, $a_{\rm p}$, to the mean free path
of the gas, $\ell$.
If $a_{\rm p}/\ell\gg1$, as is the case for dust particles in the Earth's
atmosphere, one can use the Stokes formula for the drag force,
\EQ
\FF_{\rm D}=-6\pi\rho\nu a_{\rm p}\uu_{\rm p}\quad
\mbox{(Stokes drag)},
\EN
where $\uu_{\rm p}$ is the velocity of the particle.
Note that the drag force is linear in $u_{\rm p}$, so we can write
$F_{\rm D}$ as $m_{\rm p}u_{\rm p}/\tau_{\rm s}$, where $m$ is the mass
of the particle.
To get the stopping time, this drag force must be balanced by
the mass of the particle times its acceleration, i.e.\
\EQ
m{\dd\uu_{\rm p}\over\dd t}=-\FF_{\rm D}\equiv-m{\uu_{\rm p}\over\tau_{\rm s}}.
\EN
We see that the particle speed declines indeed exponentially and that
the stopping time is given by $\tau_{\rm s}=mu_{\rm p}/F_{\rm D}$,
where $u_{\rm p}=|\uu_{\rm p}|$.

In protostellar discs the mean free path is very long, so 
$a_{\rm p}/\ell\ll1$, and one is in the so-called Epstein regime were
(Seinfeld 1986)
\EQ
\FF_{\rm D}\approx-\rho a_{\rm p}^2 c_{\rm s}\uu_{\rm p}\quad
\mbox{(Epstein drag)}.
\EN
This basically means that the mean free path in the kinetic gas formula
for the viscosity, $\nu=\onethird\ell c_{\rm s}$, is replaced by $a_{\rm p}$,
so the effective viscosity is given by $\sim a_{\rm p}c_{\rm s}$.

Finally, when the velocity of the particle is so large that the
Reynolds number with respect to the particle radius,
$a_{\rm p}u_{\rm p}/\nu$, exceeds a certain critical value,
the flow around the body is no longer laminar, but turbulent, and
so the drag force begins to depend quadratically on the velocity
of the particle, i.e.\
\EQ
\FF_{\rm D}\approx-\rho a_{\rm p}^2 u_{\rm p}\uu_{\rm p}\quad
\mbox{(turbulent drag)}.
\EN
In the turbulent regime, the viscosity of the Stokes formula
becomes essentially replaced by $a_{\rm p} u_{\rm p}$.

\begin{figure}[t!]
\resizebox{\hsize}{!}{\includegraphics[clip=true]{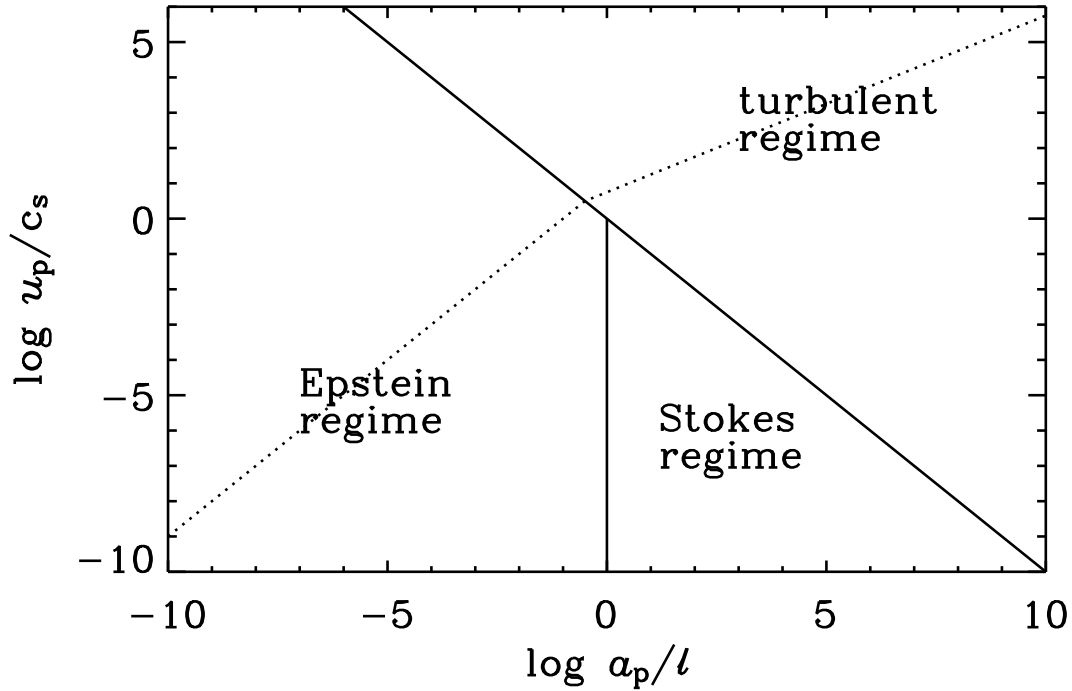}}
\caption{\footnotesize
Regimes of applicability of Stokes, Epstein, and turbulent drag
formulae as a function of particle radius, $a_{\rm p}$, and
particle speed, $u_{\rm p}$.
The dotted lines gives the terminal descent speed of particles
as a function of $a_{\rm p}$.
Note that in the turbulent regime the slope of the dotted curve
is 1/2, while in the Epstein regime it is 1.
For comparison, $\ell\approx10\m$ for the solar nebula at $1\AU$.
}
\label{pdragformula}
\end{figure}

Let us now estimate the terminal descent speed of a particle in the presence
of vertical gravity, $g=\Omega^2 z$, and let us consider a height of
one pressure scale height.
Note also that in thin discs $\Omega H/\sqrt{2}=c_{\rm s}$ is the
sound speed, but in order to avoid additional factors of $\sqrt{2}$
we prefer to use here $\tilde{H}$, so that $\Omega\tilde{H}=c_{\rm s}$.
In the following we omit the tilde.
The terminal descent speed is obtained by balancing the drag force with
the gravitational force, $mg$.
In the Epstein regime we have therefore
\EQ
u_{\rm p}={mg\over\rho a_{\rm p}^2 c_{\rm s}}\quad\mbox{(terminal speed)}.
\EN
Writing this in terms of the Mach number, $u_{\rm p}/c_{\rm s}$,
and expressing the mass in terms of radius and density of the solid,
$\rho_{\rm p}\gg\rho$, via $m\approx\rho_{\rm p}a_{\rm p}^3$, we have
\EQ
{u_{\rm p}\over c_{\rm s}}
={\rho_{\rm p}a_{\rm p}^3\over\rho a_{\rm p}^2}\,{g\over c_{\rm s}^2}
={\rho_{\rm p}\over\rho}{a_{\rm p}\over H}{z\over H}
={\rho_{\rm p}\over\rho}{a_{\rm p}\over\ell}{\ell\over H}{z\over H}.
\EN
The mean free path can be expressed in terms of viscosity by using a
formula from kinetic gas theory, $\nu={1\over3}c_{\rm s}\ell$.
Thus, $\ell/H=3\nu/(c_{\rm s}H)\approx\Rey^{-1}$.
Using $\Rey=10^9$ for the Reynolds number in protostellar discs at $1\AU$
(Brandenburg \& Subramanian 2005), together with $z\approx H$, and
$\rho_{\rm p}/\rho\approx10^{10}$ (see Hodgson \& Brandenburg 1998), we have
\EQ
{u_{\rm p}\over c_{\rm s}}\approx10\times{a_{\rm p}\over\ell}.
\EN
The dependence $u_{\rm p}/c_{\rm s}$ versus $a_{\rm p}/\ell$
is shown in \Fig{pdragformula} both for laminar flows in the
Epstein regime and for turbulent flows where the product of
$u_{\rm p}/c_{\rm s}$ and $a_{\rm p}/\ell$ exceeds unity.
The stopping time expressed in non-dimensional form is given by
\EQ
\Omega\tau_{\rm s}
=\Omega{mu_{\rm p}\over F_{\rm D}}
={\rho_{\rm p}\over\rho}{a_{\rm p}\over H}.
\EN
Using again $\rho_{\rm p}/\rho\approx10^{10}$ we find
$\Omega\tau_{\rm s}=1$ for meter-sized particles.

\begin{figure}[t!]
\resizebox{\hsize}{!}{\includegraphics[clip=true]{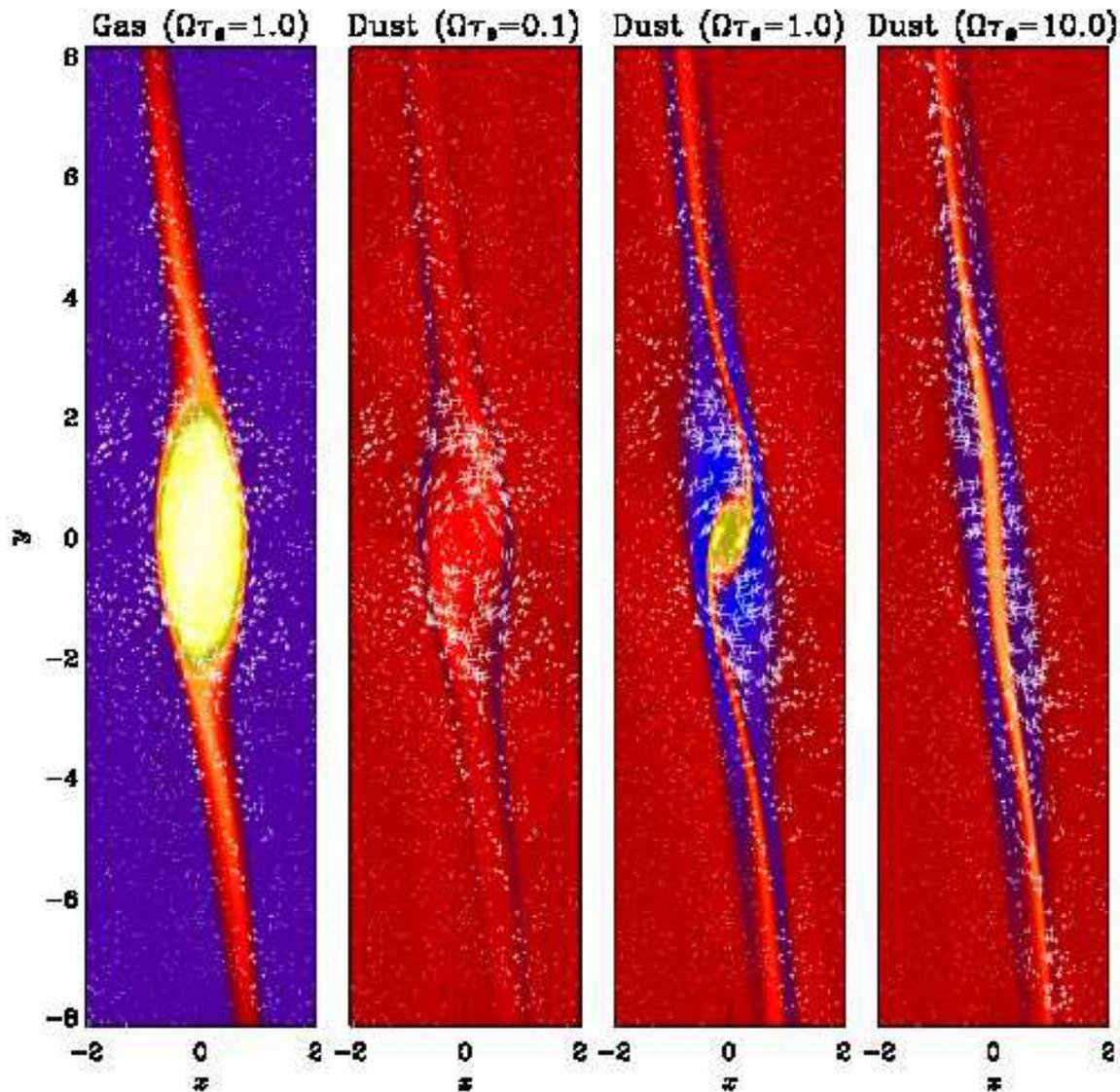}}
\caption{\footnotesize
Gas and dust in the mid-plane after one orbit.
Gas density and velocity field for the three values of
$\Omega\tau_{\rm s}$ are indistinguishable,
so only $\Omega\tau_{\rm s}=1$ is shown.
The other three plots show dust density and dust
velocity field after one orbit for $\Omega\tau_{\rm s}=0.1$,
$\Omega\tau_{\rm s}=1$ and $\Omega\tau_{\rm s}=10$ respectively.
The dust velocity field of $\Omega\tau_{\rm s}=0.1$ is very similar to
that of the gas, due to the short stopping time.
For $\Omega\tau_{\rm s}=1$ there is a strong convergence towards
the interior of the vortex,
whereas for $\Omega\tau_{\rm s}=10$ only a slight density increase
in a narrow region that extends from the vortex along the shear is seen.
Courtesy of Johansen et al.\ (2004).
}
\label{panel_xy}
\end{figure}

\begin{figure}[t!]
\resizebox{\hsize}{!}{\includegraphics[clip=true]{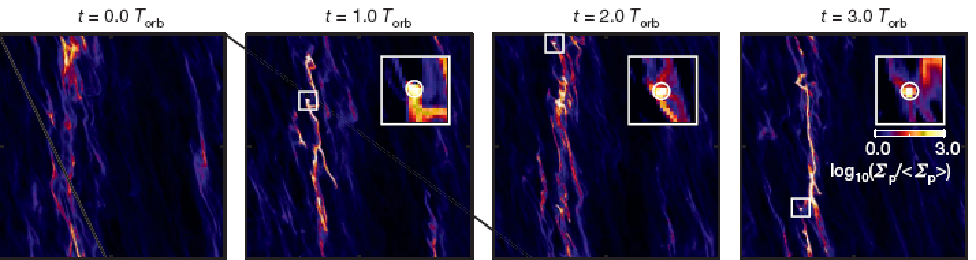}}
\caption{\footnotesize
Simulations of Johansen et al.\ (2007) showing the total column density
(gas plus all particle sizes) in the horizontal plane.
The insets show the column density in logarithmic scale centered around
the most massive cluster in the simulation.
Time is given in terms of the number of orbits after turning on self-gravity.
Courtesy of Johansen et al.\ (2007).
}
\label{johansen1}
\end{figure}

\subsection{Particle trapping in vortices}

In the following we discuss why in the size range where
$\Omega\tau_{\rm s}=1$ particles can be trapped by vortices.
This possibility was first suggested by Barge \& Sommeria (1995) and
Tanga et al.\ (1996) who proposed that dust particles
could be trapped within turbulent anticyclonic eddies.
This mechanism can be understood by taking the divergence of
the evolution equations for the particle velocity,
\EQ
{\partial\uu_{\rm p}\over\partial t}=\grav-2\OO\times\uu_{\rm p}
+3\Omega^2\xx_{\rm p}-\tau_{\rm s}^{-1}(\uu_{\rm p}-\uu),
\EN
where $\uu$ is the velocity of the gas.
Taking separately the curl and the divergence of this
equation we obtain two scalar equations,
\EQ
{\partial\over\partial t}\left(2\OO\cdot\oo_{\rm p}\right)
=-4\Omega^2\left(\nab\cdot\uu_{\rm p}\right)
-2\OO\tau_{\rm s}^{-1}\cdot(\oo_{\rm p}-\oo),
\EN
\EQ
{\partial\over\partial t}\left(\nab\cdot\uu_{\rm p}\right)
=2\OO\cdot\oo_{\rm p}+2\Omega^2
-\tau_{\rm s}^{-1}(\nab\cdot\uu_{\rm p}),
\EN
where $\oo=\nab\times\uu$ and $\oo_{\rm p}=\nab\times\uu_{\rm p}$
are the vorticity of gas and particles, respectively,
and we have assumed that the gas is solenoidal, i.e.\ $\nab\cdot\uu$,
and that only horizontal velocity components enter.
Eliminating $2\OO\cdot\oo_{\rm p}$ from these two equations yields
(Hodgson \& Brandenburg 1998)
\EQ
\left[(2\Omega\tau_{\rm s})^2+(1+\tau_{\rm s}\partial_t)^2\right]
\left(\nab\cdot\uu_{\rm p}\right)
=2\OO\tau_{\rm s}\cdot\left(\oo+\OO\right).
\EN
This equation shows that there will be a trend toward a negative
divergence (i.e.\ a positive convergence) of the particle velocity if
\EQ
2\OO\cdot(\oo+\OO)<0.
\EN
Thus, for particle accumulation, i.e.\ $\nab\cdot\uu_{\rm p}<0$,
not only does the vorticity have to be anticyclonic, but also
the vorticity has to be sufficiently anticyclonic so that
the condition above is satisfied.
Looking at histograms of vorticity, it is clear that this condition
is only satisfied in the very tail of the distribution of the
axial component of the vorticity (see Fig.~2 of Hodgson \& Brandenburg 1998).
The trapping of gas by anticyclonic vortices has been studied by a
number of people (Barge \& Sommeria 1995, Tanga et al.\ 1996,
Klahr \& Henning 1997, Chavanis 2000, Johansen et al.\ 2004).
In \Fig{panel_xy} we reproduce the result of a simulation by
Johansen et al.\ (2004), where density and velocity of the gas are
compared with density and velocity of the dust for an anticyclonic vortex
with three values of $\Omega\tau_{\rm s}$.

\subsection{Planetesimal formation with self-gravity}

A number of new simulations have emerged in recent years.
A major step was made in a paper by Johansen et al.\ (2007)
who combined the dust dynamics with self-gravity in the
shearing box approximation (Fig.~4).
One of the remarkable results they find is a rapid formation
of Ceres-sized bodies from boulders.
Even though the mass of what one might call protoplanet
is growing, this body is also shedding mass during encounters
with ambient material as it flows by.
One might speculate that what is missing is the effect of
radiative cooling of the protoplanet.
This would allow the newly accreted material to lose entropy,
become denser, and hence fall deeper into its potential well.

\section{Outflows from protostellar discs}

Finally we discuss the phenomenon of outflows and collimated
jets from protostellar discs.
A common approach to modelling jets is by treating the disc
as a boundary condition where a poloidal magnetic field is
wound up by Keplerian rotation in the disc.
A particularly useful setup has been discussed and studied
in the papers by Ouyed \& Pudritz (1997a,b) and Fendt \& Elstner (1999).

By modelling the disc as a boundary condition, it is impossible to account
for the generation and evolution of magnetic fields that are responsible
for the centrifugal acceleration of outflows via the
Blandford and Payne (1982) mechanism.
This was the reason why von Rekowski et al.\ (2003) proposed
a mean field dynamo model that allows for outflows.
We briefly summarize some of the main findings.

\subsection{Models without magnetized central star}

\begin{figure}
\centering
\includegraphics[height=6cm]{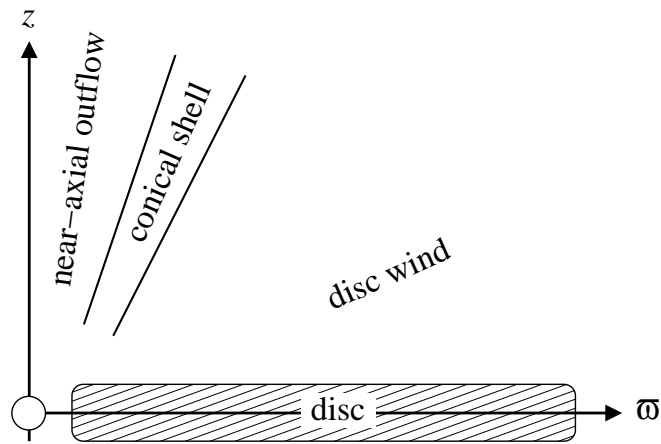}
\caption{General structure of the outflows typically obtained in
the model of von Rekowski et al.\ (2003), where the cool, dense disc emits
a thermally driven wind and a magneto-centrifugally driven outflow in a
conical shell.
[Adapted from von Rekowski et al.\ (2003).]
}
\label{Fig-structure}
\end{figure}

A mean field dynamo model with piece-wise polytropic hydrodynamics
was proposed by von Rekowski et al.\ (2003).
Like Ouyed \& Pudritz (1997a,b) they start with an equilibrium
corona, where they assume constant entropy and hydrostatic equilibrium
according to $c_p T(r)=GM_*/r$, i.e.\ enthalpy equals the
negative gravitational potential.
In order to make the disc cooler, a geometrical
region for the disc is prescribed (see \Fig{Fig-structure}).
The entropy contrast between disc and corona is chosen such that the initial
disc temperature is about $3\times10^3 \K$ in the bulk of the disc.
The low disc temperature corresponds to a high disc density of about
$10^{-10} \dots 10^{-9} \g \cmcube$.
For the disc dynamo, the most important parameter is the dynamo
coefficient $\alpha_{\rm dyn}$ in the mean field induction equation.
The dynamo $\alpha$ effect is antisymmetric about the midplane and restricted
to the disc.

\begin{figure}
 \centering
 \includegraphics[height=8.5cm]{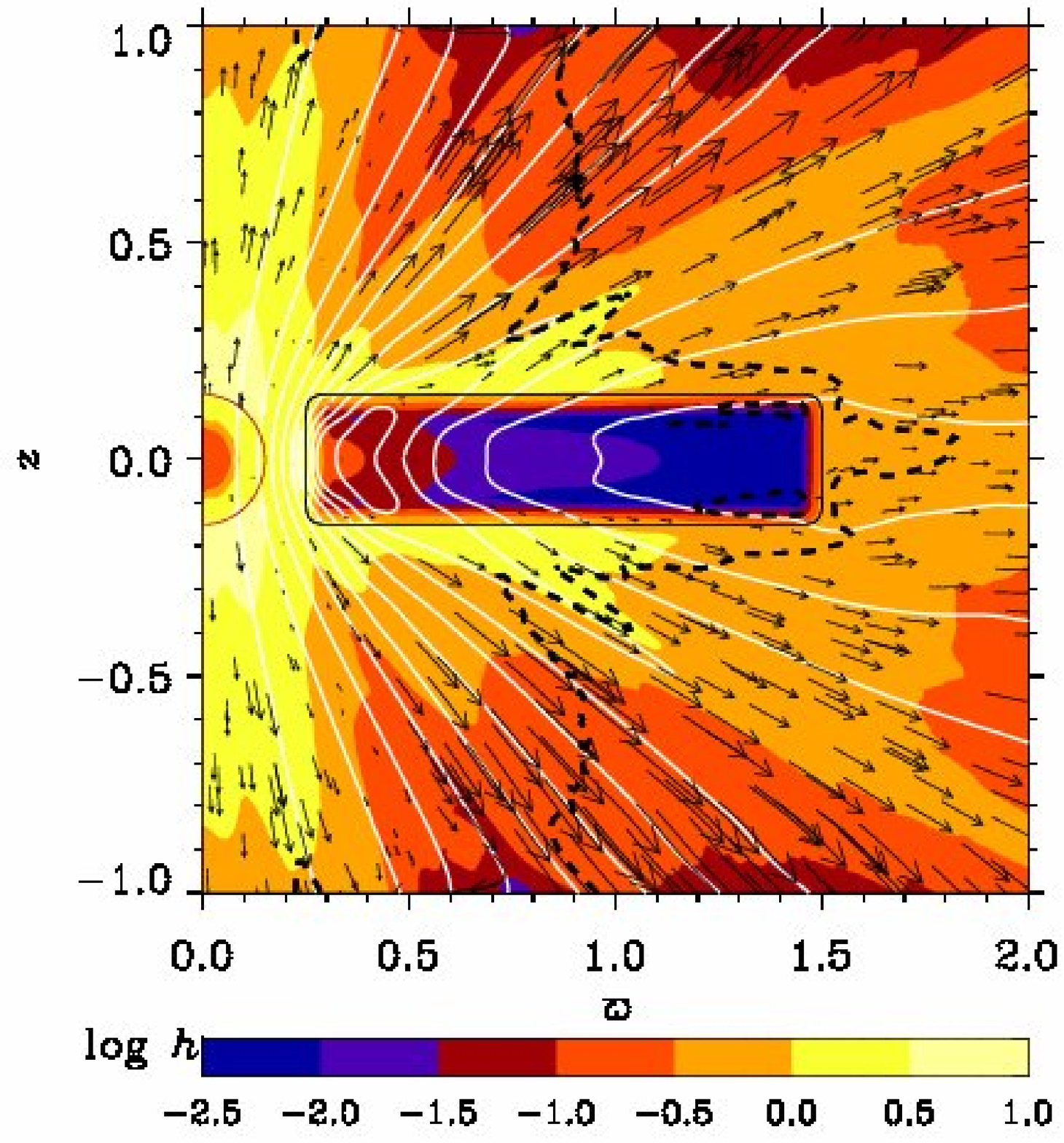}
 \includegraphics[height=8.5cm]{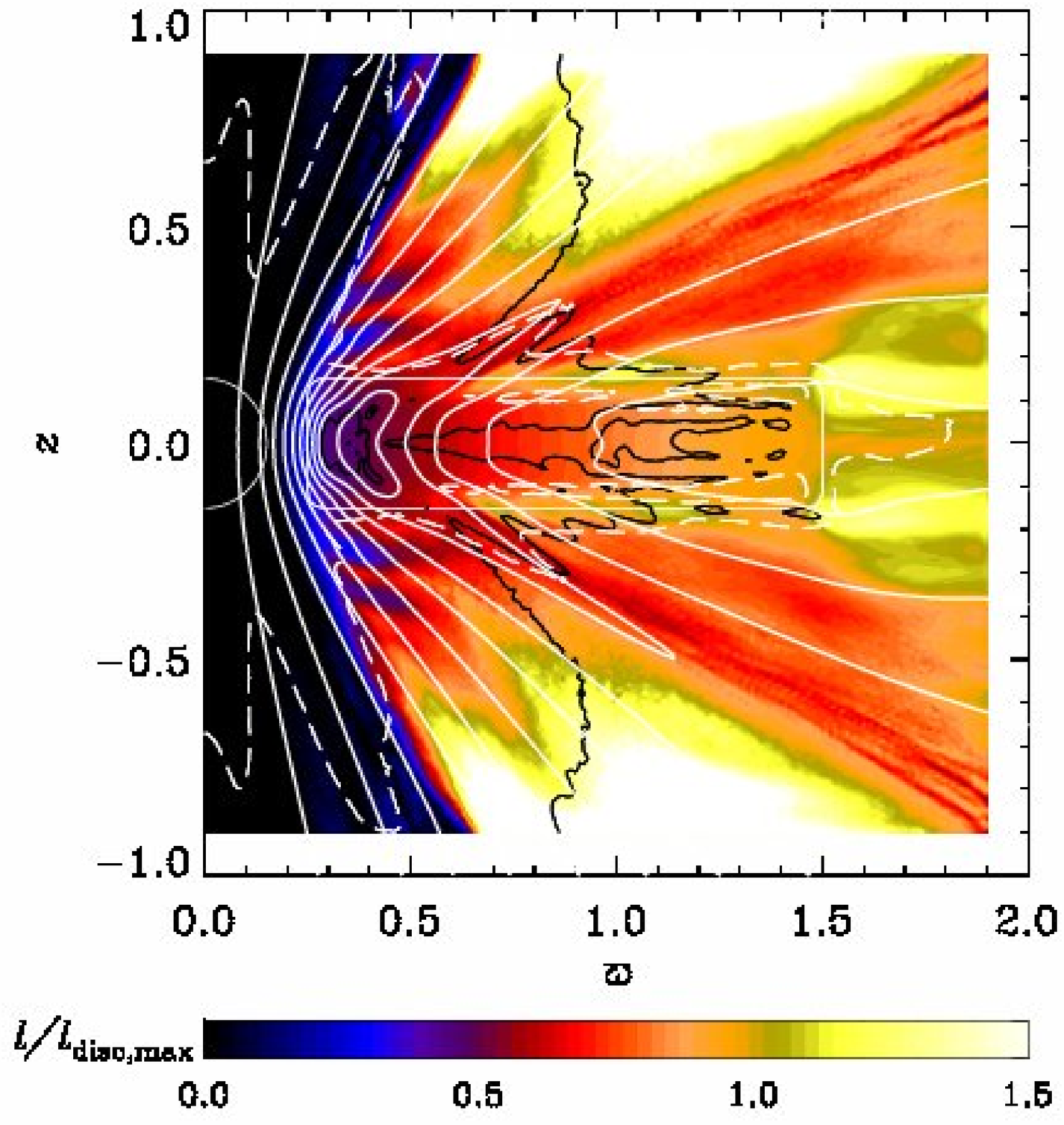}
 \caption{
  Left: 
  poloidal velocity vectors and poloidal magnetic field lines (white)
  superimposed on a color scale representation of $\log h$. Specific
  enthalpy $h$ is directly proportional to temperature $T$, and
  $\log h=(-2,-1,0,1)$ corresponds to $T\approx
  (3{\times}10^{3},3{\times}10^{4},3{\times}10^{5},3{\times}10^{6})\,\mbox{K}$.
  The black dashed line shows the fast magnetosonic surface.
  The disc boundary is shown
  with a thin black line, the stellar surface is marked in red.
  The dynamo $\alpha_{\rm dyn}$ coefficient is negative in the upper disc half,
  resulting in roughly dipolar magnetic
  symmetry.
  Right:
  color scale representation of the specific
  angular momentum, normalized by the maximum angular momentum in the disc,
  with poloidal magnetic field lines superimposed (white).
  The black solid line shows the Alfv\'en surface, the white dashed
  line the sonic surface.
  Same model as in the left hand panel and averaged over same times.
  [Adapted from von Rekowski \& Brandenburg (2004).]
 }
 \label{Fig1} 
\end{figure}

The upper panel of Fig.~\ref{Fig1} illustrates that an outflow develops
that has a well-pronounced structure. Within a conical shell originating
from the inner edge of the disc the terminal outflow speed exceeds
$500\kms$ and temperature and density are lower than elsewhere.
The inner cone around the axis is the hottest and densest region
where the stellar wind speed reaches about $150\kms$. The wind that develops
from the outer parts of the disc has intermediate values of the speed.

The structured outflow is driven by a combination of different processes.
A significant amount of angular momentum is transported outwards
from the disc into the wind along the magnetic field, especially along the
concentrated lines within the conical shell (see the lower panel of \Fig{Fig1}).
The magnetic field geometry is such that the angle between the rotation
axis and the field lines threading the disc exceeds $30^\circ$ at the disc surface,
which is favorable for magneto-centrifugal acceleration (Blandford \& Payne
1982). However, the Alfv\'en surface is so close to the disc surface
at the outer parts of the disc that acceleration there is mainly due to
the gas-pressure gradient. In the conical shell, however, the outflow is highly
supersonic and yet sub-Alfv\'enic, with the Alfv\'en radius a few times larger
than the radius at the footpoint of the field lines at the disc surface.
The lever arm of about 3 is sufficient for magneto-centrifugal acceleration
to dominate in the conical shell (cf.\ Krasnopolsky et al.\ 1999).

\subsection{Star--disc coupling}

\begin{figure}[t!]
\centering
\resizebox{.8\hsize}{!}{\includegraphics[clip=true]{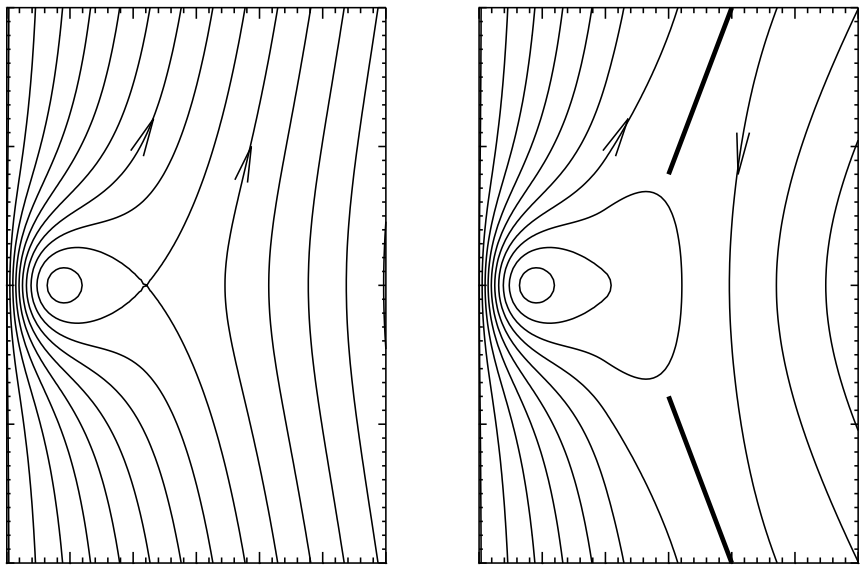}}
\caption{\footnotesize
Sketch showing the formation of an {\sf X}-point when the disc field is
aligned with the dipole (on the left) and the formation of current sheets
with no {\sf X}-point if they are anti-aligned (on the right).
The two current sheets are shown as thick lines.
In the present paper, the second of the two configurations emerges
in all our models,
i.e.\ with current sheets and no {\sf X}-point.
[Adapted from von Rekowski \& Brandenburg (2004).]
}
\label{xpoint}
\end{figure}

We now discuss
the interaction of a stellar magnetic field with a circumstellar accretion
disc and its magnetic field.
This problem was originally studied in connection with
accretion discs around neutron stars (Ghosh \& Lamb 1979),
but it was later also applied to protostellar magnetospheres
(K\"onigl 1991, Cameron \& Campbell 1993, Shu et al.\ 1994).
Most of the work is based on the assumption that the field in the
disc is constantly being dragged into the inner parts of the
disc from large radii.
The underlying idea is that a magnetized molecular cloud collapses,
in which case the field in the central star and that in the disc are aligned.
This scenario was studied numerically by Hirose et al.\ (1997) and
Miller \& Stone (1997).
In the configurations they considered, there
is an {\sf X}-point in the equatorial plane
(see left hand panel of Fig.~\ref{xpoint}),
which can lead to a strong funnel flow.

However, the opposite relative orientation is in principle also possible
and has been explored by Lovelace et al.\ (1995),
where the magnetic field of the star has been {\it flipped} and is now
anti-parallel with the field in the disc, so that the field in the
equatorial plane points in the same direction and has no {\sf X}-point.
However, current sheets develop above and below the disc plane
(see right hand panel of Fig.~\ref{xpoint}).
This configuration is also referred to as the {\sf X}-wind model.
Ironically, this is the field configuration without an {\sf X}-point.

Simulations of such a field configuration
by Hayashi et al.\ (1996) confirm the idea by Lovelace et al.\ (1995)
that closed magnetic loops connecting the star with disc are twisted
by differential rotation between star and disc, and then inflate
to form open stellar and disc field lines (see also Bardou 1999,
Agapitou \& Papaloizou 2000, Uzdensky et al.\ 2002).
Goodson et al.\ (1997, 1999) and Goodson \& Winglee (1999) find that
for sufficiently low resistivity, an accretion process develops that is
unsteady and proceeds in an oscillatory fashion.
Similar results have also been obtained by von Rekowski \& Brandenburg (2004).
In their case the resulting field geometry
is always the one in the second panel of Fig.~\ref{xpoint},
i.e.\ the one with current sheets and no {\sf X}-point.
It is seen that the inflating magnetosphere expands to
larger radii where matter can be loaded onto the field lines and be ejected as
stellar and disc winds.
Furthermore, reconnection of magnetic field lines allows matter
to flow along them and accrete onto the protostar, in the form of a funnel flow
(Romanova et al.\ 2002).

\section{Conclusions}

Protostellar discs are generally believed to be turbulent, although
there is substantial uncertainty regarding the relative importance of
possible mechanisms and the necessary physics involved.
The turbulence in protostellar discs may not be as vigorous as in fully
ionized discs around neutron stars and black holes, but its presence
is critical for driving accretion toward the central star,
driving mass concentrations within the disc above the critical value
for gravitational collapse to form planets.
Turbulence is also important for heating a corona
above the disc from where outflows can be driven.

Several important aspects have been omitted in this review.
The inclusion of radiative cooling has already been mentioned as one
of the important ingredients that allow matter to settle deeper in the
potential well and hence to prevent newly accreted material from being
stripped away from the forming protoplanet.
Related to this is the question how effective the acceleration of outflows
is when one allows for radiative cooling of the jet.
Another aspect concerns the generation of large scale magnetic fields both
in the protostar and the disc.
Much attention has been paid to the importance of shedding small scale
magnetic helicity from the dynamo, because otherwise the dynamo will
suffocate from excess small scale magnetic helicity that quenches
the production of large scale magnetic fields by kinetic helicity
and/or shear.
Advanced global simulations are likely to shed light on these and
related questions.

\section*{References}

\bibliographystyle{aa}

\end{document}